\documentclass[%
 reprint,
 amsmath,amssymb,
 aps,
]{revtex4-2}
\usepackage{amsmath} 
\usepackage{graphicx}% Include figure files
\usepackage{dcolumn}% Align table columns on decimal point
\usepackage{bm}% bold math
\usepackage{color}
\begin{document}

\preprint{APS/123-QED}

\title{Driven Odd Elasticity in Passive Mechanical Metamaterials}% Force line breaks with \\

\author{Mohamad Rahimi}
 \altaffiliation{Department of Mechanical Engineering, Boston University, 02215, Boston, MA, USA}
\author{Harold S Park}%
 \email{parkhs@bu.edu}
\affiliation{%
Department of Mechanical Engineering, Boston University, 02215, Boston, MA, USA
}%

\begin{abstract}
We present a mechanical mechanism leveraging passive mechanical components, i.e. chiral gears and a square lattice metamaterial, to demonstrate driven odd elasticity in a mechanical metamaterial.  The mechanism couples tension and shear in a non-reciprocal way, resulting in an odd shear modulus. The emergence of this odd shear modulus enables non-conservative work in a standard quasistatic strain cycle, and further enables the non-Hermitian skin effect in dynamics. Our results demonstrate that odd elasticity can be achieved in mechanical structures using passive elements without electronic components coupled with feedback or robotic control systems.
\end{abstract}

%\keywords{Suggested keywords}%Use showkeys class option if keyword
                              %display desired
\maketitle

%\tableofcontents

The behavior of a linear elastic isotropic solid can be described by a free energy function, and therefore such solids conserve linear and angular momentum and energy, and are mechanically passive, i.e. they cannot do work on their surroundings \cite{fruchart2023odd}.  Furthermore, the constitutive relations that describe the stress-strain relationships of such solids are symmetric, or reciprocal, and satisfy the well-known Maxwell-Betti reciprocity~\cite{coulais2017static, achenbach2003reciprocity,montazeri2025non}.  Recently, the concept of an odd elasticity was theoretically introduced~\cite{scheibnerNP2020} to describe linear elastic isotropic solids whose mechanical behavior cannot be described by a free energy function.  As a result, odd elastic solids have a non-symmetric elasticity tensor that captures nonreciprocity, where the mechanical response to different loads is not the same.  For example, in an odd elastic solid, extension could induce torque, while the same torque would not induce extension~\cite{scheibnerNP2020}.  A 2D odd elastic solid is also chiral, so it cannot be superimposed on its mirror transformation through any translation or rotation, which enables the coupling of different deformation fields (i.e. tension to shear) that would otherwise require anisotropy to achieve~\cite{liuJMPS2012,fernandezAM2019,liuJME2016,wuMD2019}.  Most importantly, while an odd elastic solid would conserve linear momentum, it would, intriguingly, not be required to conserve angular momentum or energy.  Odd elasticity has been widely studied for a range of applications, including self-locomotion of robots~\cite{veenstra2025adaptive}, localizing, attenuating and amplifying elastic waves in a non-Hermitian fashion~\cite{scheibnerPRL2020,chenNC2021}, biological systems \cite{shankarNP2024,banerjee2021active}, active matters \cite{markovich2021odd,tan2022odd,chao2026selective}, and odd fluids \cite{duclut2024probe}.

The above theory and applications of odd elasticity have assumed, for theoretical studies~\cite{scheibnerNP2020,scheibnerPRL2020,shaat2025non,wu2023active,wu2023engineering,chengSC2021,lai2024odd,zhou2020non} and both non-biological~\cite{brandenbourgerNC2019,chenNC2021,veenstra2025adaptive} and biological experimental studies~\cite{tanNATURE2022,shankarNP2024}, the presence of an internal energy source, or electronic components coupled with feedback or robotic control~\cite{brandenbourgerNC2019,chenNC2021,veenstra2025adaptive} to enable the odd elastic behavior, and as such are considered variants of ``active" odd elasticity.  In contrast, the notion of driven odd elasticity was recently introduced~\cite{huangARXIV2023}, where the application of a periodic (driving) mechanical loading served as the input energy source to activate a theoretical non-reciprocal frictional response to enable a structure to exhibit an odd elastic response.  This notion of driven odd elasticity is promising because external mechanical loading at the boundaries is the standard approach to applying forces on structures, and as such has the potential to enable odd elasticity in mechanical structures.  However, exploiting driven odd elasticity still has an open challenge: the development of an internal mechanism, that does not require an internal energy source or electronic components, that causes a structure to respond non-reciprocally to applied loads.  As such, to our knowledge, neither driven odd elasticity~\cite{huangARXIV2023} nor the originally proposed odd elasticity~\cite{scheibnerNP2020} have been demonstrated in a mechanical system using only passive components and without electronic control or feedback systems.

Here, we demonstrate driven odd elasticity using passive components:  a square lattice metamaterial, and chiral gears.  Unlike known odd elastic media~\cite{chenNC2021,tanNATURE2022,shankarNP2024,veenstra2025adaptive} that have an internal source of energy (for biological active matter)~\cite{tanNATURE2022,shankarNP2024} or electronic components to achieve local feedback control~\cite{chenNC2021,veenstra2025adaptive}, we use driven chiral gears that act on the boundaries of the lattice metamaterial to achieve a non-reciprocal coupling between tension and shear. This results in the structure exhibiting an odd shear, whereby the application of normal strain results in shear strain, but the application of shear strain does not result in normal strain.  This non-reciprocal coupling results in an asymmetric elasticity tensor, and we further demonstrate that the metamaterial reveals non-conservative work through a closed cycle of deformation. Finally, we observe that the non-reciprocal coupling between normal and shear deformation results in an effective anisotropic behavior of the system, which breaks parity-time (PT) symmetry and enables the emergence of the non-Hermitian skin effect~\cite{scheibnerPRL2020}.

\subsection{\label{sec:level1}Metamaterial design}

\begin{figure}
    \centering
    \includegraphics[width=1\linewidth]{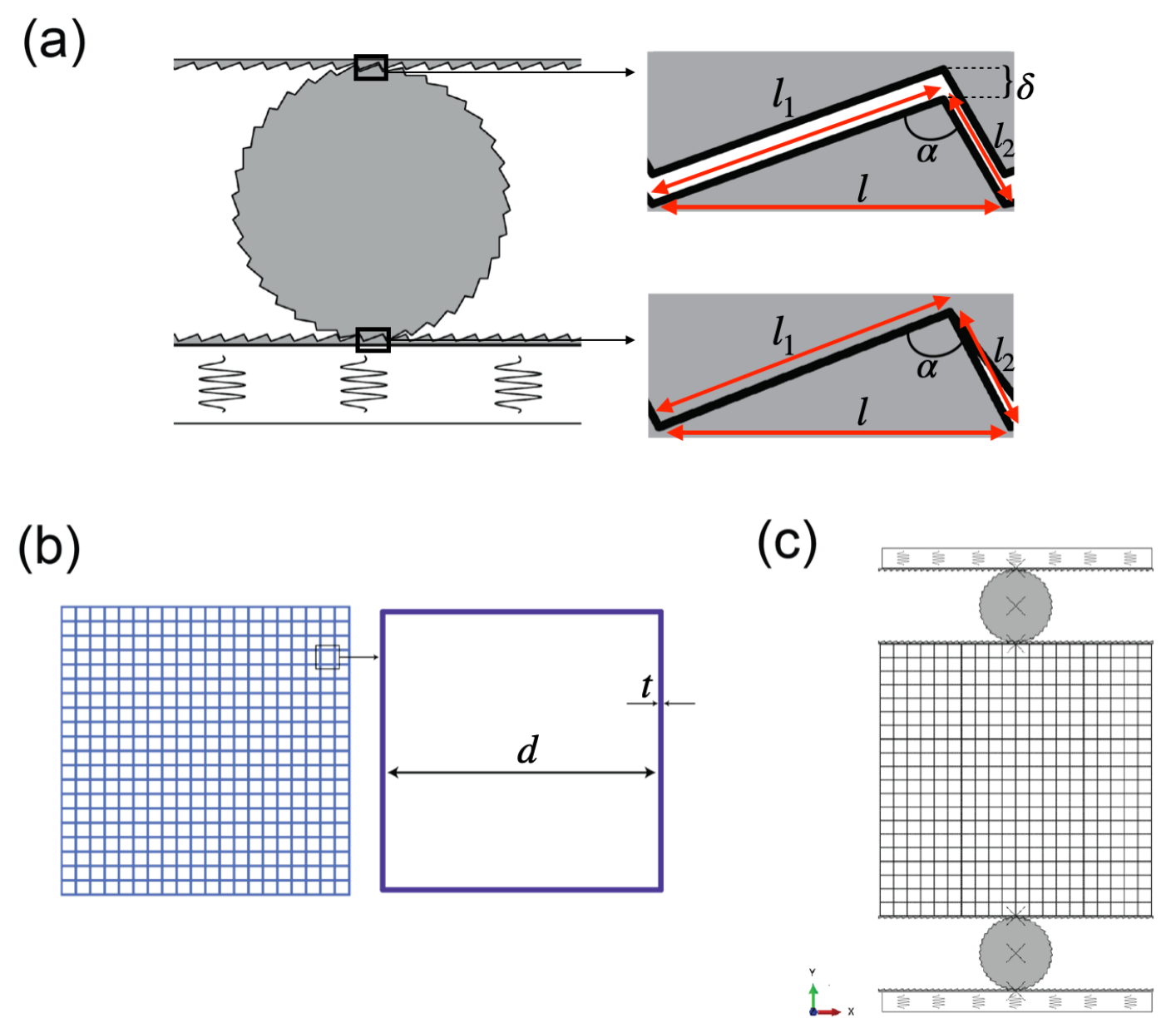}
    \caption{(a) Chiral gears on springs, with geometry of gear contacts shown.  (b) Square lattice metamaterial.  (c) Coupled chiral gear plus metamaterial structure that exhibits driven odd elasticity.}
    \label{fig:fig1}
\end{figure}

We first introduce the components comprising the metamaterial.  The first component, as shown in Fig. \ref{fig:fig1}(a), is a gear with chiral teeth that is in contact with two horizontal walls. The gear, as well as the walls, have specific degrees of freedom described below but are not deformable.  The gear is already in contact with the bottom wall, while having a clearance of amount $\delta$ with the top wall.  All surfaces in contact are assumed to have a coefficient of friction of $\mu=0.1$, and long ($l_{1}$) and short ($l_{2}$) contact lengths.  The top wall has a horizontal degree of freedom, while the bottom wall lies on a foundation with spring stiffness of $k$= \(5\times 10^6\) N/m that enables the bottom wall to move vertically, whereas the gear can both rotate and translate vertically.  The specific dimensions for the results that follow are clearance $\delta$=0.15 mm, teeth length $l$=3.5 mm and height $h$=1 mm.

%\hsp{Should discuss here dependence of odd shear on $\mu$, and show in the SI a plot of the odd shear constant as a function of $\mu$ for $mu$ between 0 and 1.}  \mr{I think we discussed this before. For this spring stiffness the maximum friction is 0.1, otherwise the spring should be softer.}

The second component of the metamaterial is a square lattice with thickness $t=0.5$ mm and unit cell length $d$=16mm as shown in Fig. \ref{fig:fig1}(b), where each unit cell can stretch, shear and bend.  We choose the mechanical properties of the metamaterial to be a Young's modulus of $E$=300 GPa, a Poisson's ratio of $\nu$=0.3, and a density of $\rho$=8000 kg/m$^{3}$, to represent a generic class of metallic materials.  As shown in Fig. \ref{fig:fig1}(c), we then attach the gear mechanism in Fig. \ref{fig:fig1}(a) to the bottom of the lattice metamaterial in Fig. \ref{fig:fig1}(c), while the same gear structure in (a) is attached to the top surface of the lattice metamaterial by flipping it upside down in Fig. \ref{fig:fig1}(c).  The coupled mechanism plus metamaterial structure in Fig. \ref{fig:fig1}(c) can then be driven by applying oscillatory motion of the gears.  The importance of the gears having a chiral nature for the coupled gear plus lattice metamaterial to exhibit odd elasticity will be demonstrated next.

\subsection{\label{sec:level1}Mechanism for driven odd elasticity}

\begin{figure*}
\includegraphics[width=.9\linewidth]{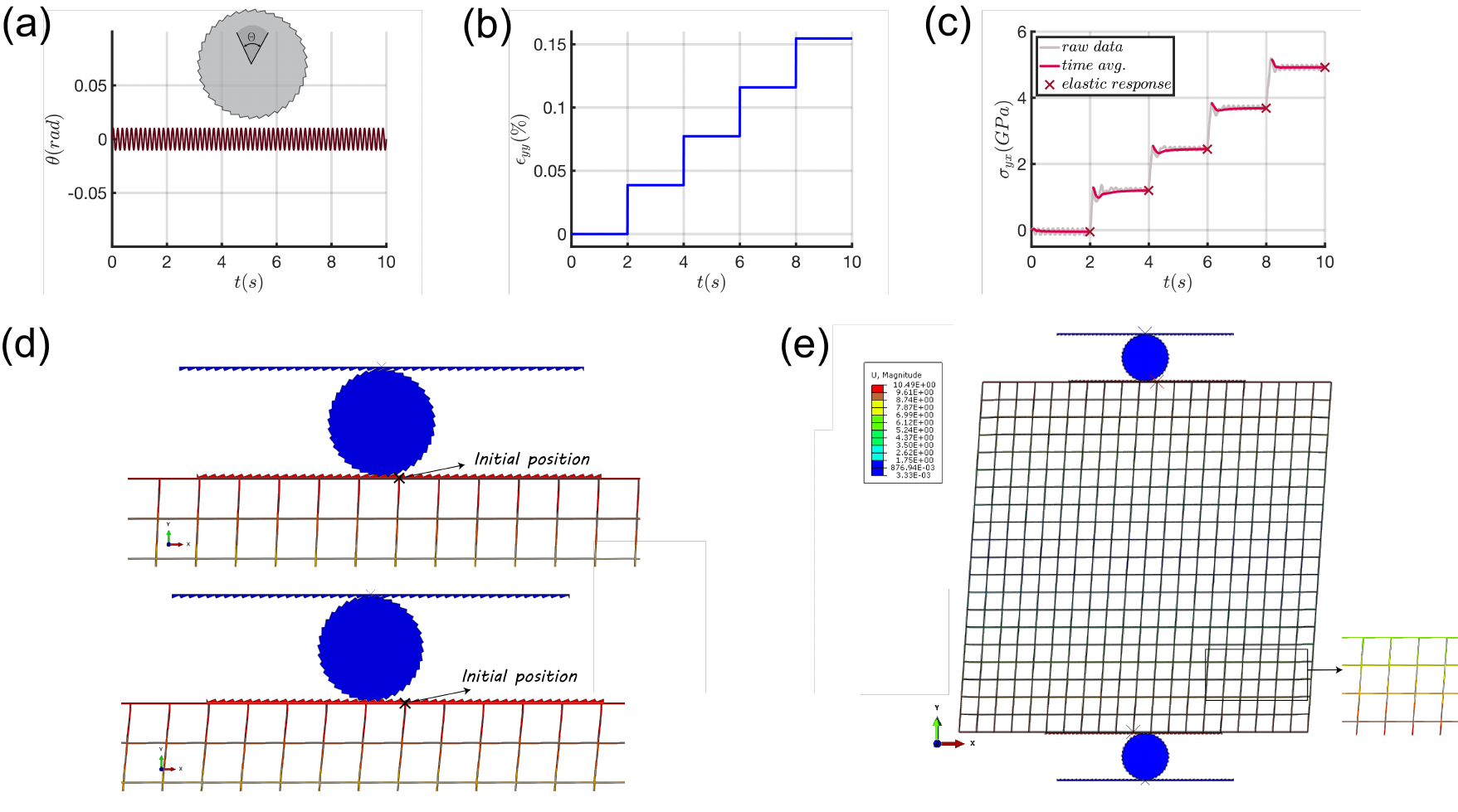}% Here is how to import EPS art
\caption{\label{fig:fig2}(a) Illustration of periodic (driven) oscillations applied to each gear. (b) Normal strain increments $\epsilon_{yy}$ applied to the metamaterial.  (c) Odd shear stress $\sigma_{yx}$ resulting from each increment of normal strain $\epsilon_{yy}$ in (b). (d) The gears being displaced after two teeth (top), and four teeth (bottom) from its initial position. (e) Deformed configuration of the metamaterial after all normal strain in (b) has been applied demonstrating significant amount of odd shear deformation. The material properties are $E$=300 GPa and $\nu$=0.3.}
\end{figure*}

We now demonstrate that the coupled gear plus metamaterial structure in Fig. \ref{fig:fig1}(c) exhibits driven odd elasticity through finite element simulations using the commercial finite element software ABAQUS \cite{0b112d0e5eba4b7f9768cfe1d818872e}.  Within these simulations, we rotate (drive) the gears at a frequency of $\omega_{d}=12\pi$, with amplitude $\Theta=0.01$ radians, as shown in Fig. \ref{fig:fig2}(a).  This driven periodic oscillatory motion is analogous to the microscopic motions considered in the driven odd elasticity paper by Huang et al.~\cite{huangARXIV2023}, and we note that it could be achieved in both mechanical or non-mechanical means, for example through the torque generated by fluid flow interacting with rigid bodies~\cite{mandujano2024chaotic,bonheure2025forced}, or by the application of magnetic fields on bodies containing conductive materials~\cite{polczynski2024dynamics}.  Because of the initial contact between the gears and the walls on the elastic foundation, and because the top wall of the gear mechanism in Fig. \ref{fig:fig1}(a) is tied to both the top and bottom surfaces of the lattice metamaterial as shown in Fig. \ref{fig:fig1}(c), periodic rotation of the gears results in periodic shear forces being exerted on the metamaterial boundaries where the gears are attached.  
%Due to the geometric shape of the chiral teeth, the stiffness in the forward and backward directions is different and is defined as \(K^\pm = k(1 \pm \epsilon)\), which results in the shear strain non-reciprocal, or greater one direction than the other.  \hsp{Is this actually true? - this would mean that if you kept oscillating the gear, you would see a non-zero drift in the shear strain without needing the applied normal displacements.  Do you see that?}  

%We intend to couple axial and shear deformation to obtain odd elastic mechanism. We rotate the gears at the frequency \(\omega_d=12\pi\) with an amplitude \(\Theta=0.01 \ (rad)\) (Fig.2 (a)). Because of the initial contact between the gears and the bottom walls, periodic rotation of the gears exerts force on the walls. Consequently, this force imposes periodic horizontal displacement on the boundaries of the lattice mimicking periodic shear strain. In order to main the contact state, we restrict the rotational degree of freedom of the lattice. As the walls are tied to the boundaries of the lattice, this periodic displacement causes periodic shear strain in the lattice. Due to the geometric shape of the chiral teeth, the stiffness in the forward and backward motions is different and defined as \(K^\pm = k(1 \pm \epsilon)\). Therefore, shear strain is greater one way than the other. 

While the driving periodic oscillatory motions are applied continuously to the gears, we also apply increments of normal strain $\Delta\epsilon_{yy}=0.04\%$ on the lattice as illustrated in Fig. \ref{fig:fig2}(b).  Importantly, when each increment of normal strain is applied, the gears momentarily lose contact with the metamaterial surfaces and are detached, as shown in Video 1.  The clearance $\delta$ in Fig. \ref{fig:fig1}(a) is critical here to enable space for the gear to detach as the normal strain is applied to the lattice.  Once the gears regain contact with the walls, they do so at a new contact point, i.e. the \emph{next} gear tooth position on the contacting horizontal wall.  Because of the geometry of the horizontal walls in Fig. \ref{fig:fig1}(a) (i.e. periodic repetition of gear teeth), this jumping of the gear from one tooth to the next locks in a new equilibrium position, and results in an increment of odd shear strain due to the application of the normal strain, and as such, the development of an odd shear stress.  This is shown in Fig. \ref{fig:fig2}(c), where we plot the shear stress as a function of time. There, we can see that each application of normal strain in Fig. \ref{fig:fig2}(b) results in a corresponding increase in shear stress in Fig.  \ref{fig:fig2}(c).

Once we have applied each increment of normal strain, we allow the system to equilibrate at its new configuration (with both nonzero normal and shear strains) through continued application of periodic oscillatory motion of the gears, until the next application of normal strain.  As can be seen in Fig. \ref{fig:fig2}(c), the odd shear stress equilibrates about an equilibrium position before the next increment of normal strain is applied.  Fig.2(d) shows the deformed metamaterial for different increments of normal strain, which shows that the position of the gears relative to the metamaterial boundaries shifts for each increment of normal strain relative to the undeformed state in Fig. \ref{fig:fig1}(c), demonstrating the odd shear that has been generated through the application of the normal strain.

The chiral geometry of the gear teeth is important here, for several reasons.  First, the asymmetric contact lengths $l_{1}$ and $l_{2}$ that are present in Fig. \ref{fig:fig1}(a) mean that as the normal strain is applied, the lattice will be sheared in only one direction (i.e. in the direction normal to the larger contact length $l_{1}$).  Second, as shown in the SI, larger normal strain increments are required to generate odd shear if the gear teeth are non-chiral (symmetric), both of which motivate the usage of the chiral gear geometry.  In contrast, if the gear geometry is symmetric, then depending on the phase of the gear with respect to the wall that is tied to the metamaterial, the gear jump from the application of normal strain could result in the lattice being sheared in either direction, thus potentially cancelling any odd shear developed previously.  The phase of the gear with respect to the wall tied to the metamaterial is also relevant for chiral gears.  Specifically, because the chirality biases the lattice to be sheared in only one direction, application of normal strain can only result in either no jump (and no odd shear strain), or odd shear occurring in only one direction, depending on the phase of the gear with respect to the contact surface.  The robustness of this mechanism with respect to the various geometric and material properties mentioned above for the metamaterial in Fig. \ref{fig:fig1}(c) is discussed in detail in the SI.

%To couple modes of deformation, as the gears rotate, we apply normal strain \(\epsilon_{yy}\) instantaneously on the lattice (Fig.2 (b)). When the strain field is applied, the gears momentarily lose contact with the walls and are detached (See Video1). Due to the geometry of the biased teeth, as the strain increment is applied, the lattice is horizontally displaced. The gears retrieve contact with walls but at a new configuration. Since the walls have teeth, the gears are locked at the new contact point. Then, we let the gears rotate to find their equilibrium position. We observe that the gears equilibrate around a new well-defined reference state. Therefore, normal strain makes the gears jump over the wall teeth. \(\Delta\epsilon_{yy}=0.04\%\) is applied incrementally, and at each increment, the gears jump over one tooth. The amplitude of the rotation remains the same, while Fig.2 (c) indicates the magnitude of shear stress has changed after applying normal strain. 

Because the driving shear stress is periodic, to characterize the mechanical properties of the system, we study the time-averaged response of the system to the quasi-static normal strain $\epsilon_{yy}$ that occurs on the time scale \(t=2s\) as shown in Fig. \ref{fig:fig2}(b), which is longer than the driving period \(T = 2\pi/\omega\). As seen in Fig. \ref{fig:fig2}(c), at each period, we measure the time-averaged shear stress, which converges to a constant after a few periods. This constant is the elastic response of the system, which we use to calculate the elasticity tensor \(C_{ijkl}\), which relates the time-averaged stress \(\sigma_{ij}\) to strain \(\epsilon_{ij}\).  The constitutive equation of this model is the following
\begin{equation}\label{eq:eq1}
\begin{bmatrix}
\sigma_{xx}\\
\sigma_{yy}\\
\sigma_{yx}\\
\sigma_{xy}
\end{bmatrix}
=\mathbf{C}
\begin{bmatrix}
\epsilon_{xx}\\
\epsilon_{yy}\\
\epsilon_{yx}\\
\epsilon_{xy}
\end{bmatrix}
\end{equation}
where we write the 2D elasticity tensor as
\begin{equation}
\mathbf{C}
=
C^{e}+C^{o}
=
\begin{bmatrix}
B+G & B-G & 0 & 0\\
B-G & B+G & 0 & 0\\
0 & 0 & G & G\\
0 & 0 & G & G
\end{bmatrix}
+\begin{bmatrix}
0 & 0 & 0 & 0\\
0 & 0 & 0 & 0\\
0 & A & 0 & 0\\
0 & 0 & 0 & 0
\end{bmatrix}
\label{eq2}
\end{equation}
In Eq. (\ref{eq2}), \(B\) and \(G\) are the conventional bulk and shear moduli which are 214.3 and 115.4 GPa, respectively. However, \(A\) is an odd modulus coupling normal strain to shear stress. The elasticity matrix has symmetric and asymmetric (odd) parts \(C=C^e+C^o\), where the symmetric part $C^{e}$ is isotropic, whereas the asymmetric part $C^{o}$ indicates normal strain causes shear stress, which establishes non-reciprocity in the system.  We compute \(A\) as
\begin{equation}\label{eq:eq2}
A = \frac{\Delta \sigma_{yx}}{\Delta\epsilon_{yy}}
\end{equation}
Using Eq. (\ref{eq:eq2}), we obtain $A=3125$ GPa, which shows that a small increment of normal strain results in a significant odd shear stress. The amount of shear in each step is proportional to the length of the gear teeth $l$ that are in contact (Fig. \ref{fig:fig1} (a)).  This amount can be modified by geometric modifications, which is shown in the SI.

\begin{figure}
    \centering
    \includegraphics[width=1\linewidth]{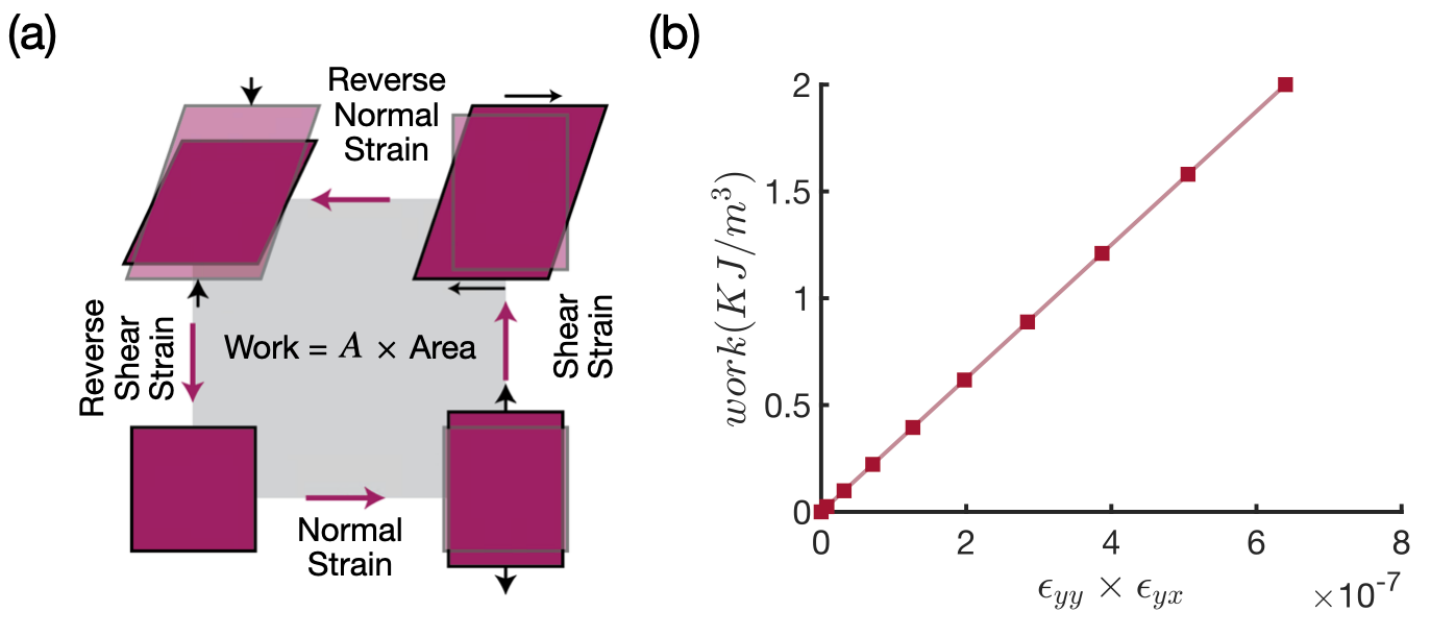}
    \caption{Illustration of non-conservative work enabled by the odd elastic metamaterial in Fig. \ref{fig:fig1}(c).}
    \label{fig:fig3}
\end{figure}

\subsection{\label{sec:statics}Statics}

One important implication of an asymmetric (odd elastic) elasticity tensor is that a structure exhibiting that mechanical response can demonstrate non-conservative work through a cycle of deformations.  We demonstrate this analytically by subjecting the metamaterial to a closed strain cycle of normal (\(\epsilon_{yy}\)) and shear (\(\epsilon_{yx}\)) strains according to Fig.\ref{fig:fig3}(a). The constitutive equations along this deformation path are
\begin{equation}\label{eq:eq3}
\begin{pmatrix}
  \sigma_{yy} \\ 
  \sigma_{yx} \\
\end{pmatrix} =  
\begin{bmatrix}
B+G & 0   \\A & G 
\end{bmatrix}
\begin{pmatrix}
  \epsilon_{yy} \\ 
  \epsilon_{yx} \\
\end{pmatrix}
\end{equation}
The net work done in a closed deformation cycle is proportional to
\begin{equation}\label{eq:eq4}
\delta w =\sigma_{ij}{\delta\epsilon_{ij}}
\end{equation}
For conventional elastic media, Eq. \ref{eq:eq4} is conservative, but this is not the case for odd elastic media, for which the work calculated through Eq. \ref{eq:eq4} has both conservative and non-conservative contributions, which can be written as
\begin{equation} \label{eq:eq5}
\begin{split}
\delta w_p & = \frac{1}{2}\delta[(B+G)\epsilon_{yy}^2+G\epsilon_{yx}^2] \\
\delta w_{np} &=A\epsilon_{yy}\delta\epsilon_{yx}
\end{split}
\end{equation}
The potential part $\delta w_{p}$ is path-independent and vanishes in a closed deformation path. On the other hand, the non-potential part $\delta w_{np}$ is path-dependent and remains non-zero at the end of the cycle, which leads to non-conservative work. Fig. \ref{fig:fig3}(b) indicates that the work at the end of the cycle is proportional to the odd modulus $A$ multiplied by the area of the strain cycle.

\subsection{\label{sec:level1}Dynamics}

\begin{figure}
    \centering
    \includegraphics[width=1\linewidth]{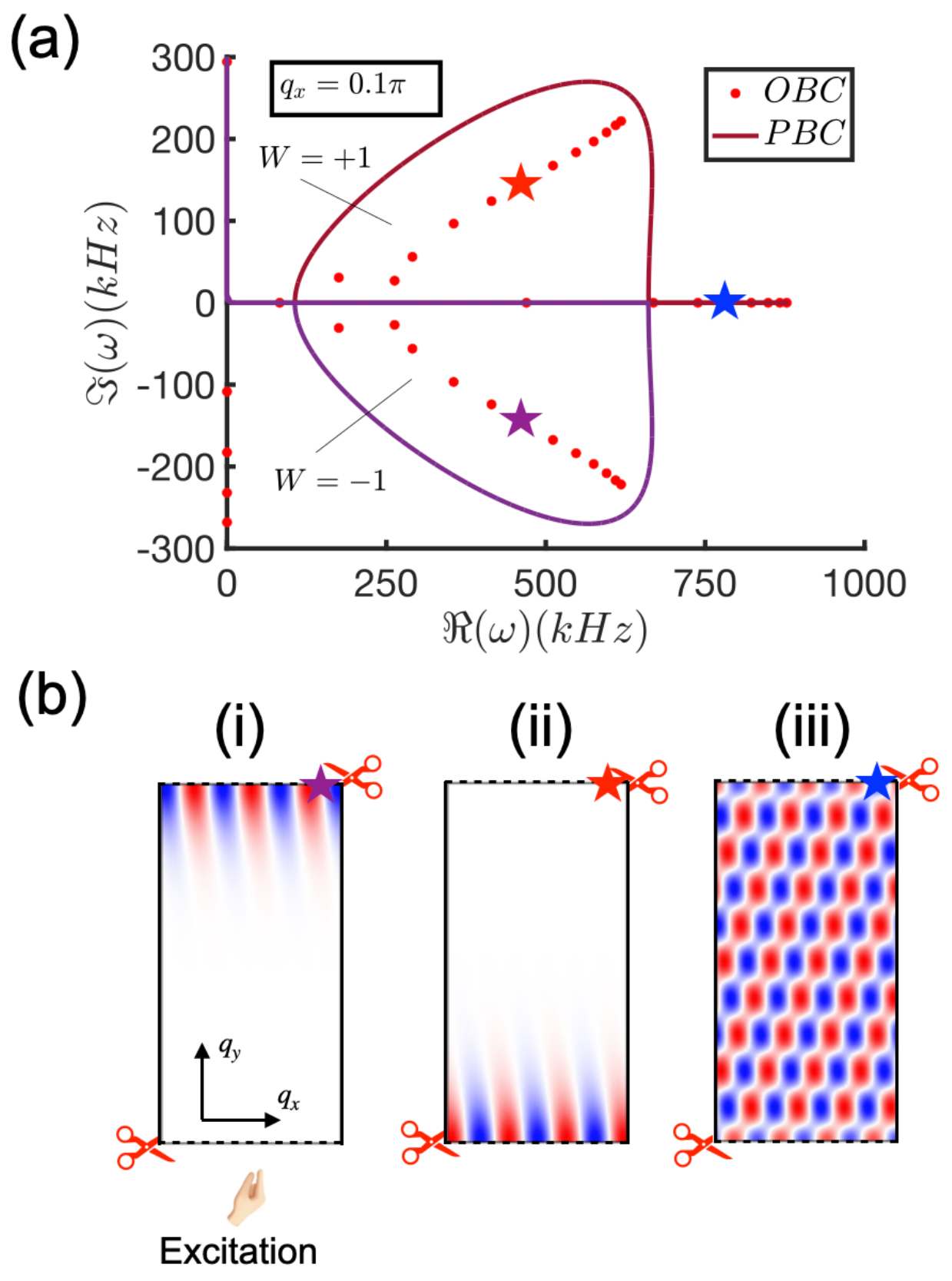}
    \caption{(a) Spectrum for odd elastic metamaterial for both open and periodic boundary conditions.  (b) Illustration of (i) amplified, (ii) attenuated, and (iii bulk) waves.  }
    \label{fig:fig4}
\end{figure}

Odd elastic solids can also exhibit interesting dynamic properties, such as the non-Hermitian skin effect (NHSE), where due to the non-reciprocal constitutive relationships, modes will localize at the boundaries of the structure when open boundary conditions are enforced~\cite{scheibnerPRL2020}.  To consider this possibility in the driven metamaterial from Fig. \ref{fig:fig1}(c), we derive the governing equations of motion. In odd elastic media, the linear momentum is conserved, and the forces are proportional to the divergence of the stress tensor.
\begin{equation}\label{eq:eq6}
\sigma_{ji,j} =\rho\ddot{u_i}
\end{equation}
In Eq. \ref{eq:eq6}, \(\ddot{( \ )}\) represents a double derivative with respect to time, and \(\rho\) and \(u_j\) are the density and displacement, respectively. Substituting the constitutive relations in Eq. \ref{eq:eq1} into Eq. \ref{eq:eq6} leads to the two equations of motion that govern the dynamics of the system.
\begin{equation} \label{eq:eq7}
\begin{split}
    (B+G)\frac{\partial^2 u_x}{\partial x^2} + B\frac{\partial^2 u_y}{\partial x \partial y} +A \frac{\partial^2 u_y}{\partial y^2} + G \frac{\partial^2 u_x}{\partial y^2}=\rho \ddot{u_x} \\
    (B+G)\frac{\partial^2 u_y}{\partial y^2} + B\frac{\partial^2 u_x}{\partial x \partial y}+G \frac{\partial^2 u_y}{\partial x^2} =\rho \ddot{u_y}
\end{split}
\end{equation}
The system has two degrees of freedom \(u_x\) and \(u_y\) and thus two equations of motion, where the odd modulus $A$ appears in the second equation for $\ddot{u}_{y}$.  We assume harmonic wave solutions \(u_j=\hat{U_j}e^{i(q.x-\omega t)}\) for the discretized model of Eq. \ref{eq:eq7}, where \(q\) is the wave vector. The dynamical matrix of Eq. \ref{eq:eq7} is non-Hermitian because of the odd modulus $A$, and this presence of $A$ thus breaks PT symmetry in the \(x\) or \(y\) directions separately, implying the localization of eigenmodes at the edges. To find the eigenfrequencies of the system, we fix \(q_x\) and assume periodic boundary conditions in the \(y\)-direction. Due to the symmetry in the \(x-\)direction, we consider the Irreducible Brillouin Zone (IBZ), for which the spectrum comes in complex conjugate pairs for all \(0 < q_x<\pi\). The eigenspectra of the periodic system are plotted in Fig. \ref{fig:fig4}. It is observed that the frequency forms two closed bands that indicate the frequency in these regions is complex conjugate. For all the eigenfrequencies in these two bands, the corresponding eigenmodes are localized at the top and bottom edges of the metamaterial. Using the definition of the winding number, we can find the direction of the localization.
\begin{equation}
W(\omega_0) = \frac{1}{2\pi i} \sum_{\alpha} \oint_{-\pi}^{\pi} \frac{d}{dq} \log[\omega_{\alpha}(q) - \omega_0] \, dq
\end{equation}
For every frequency in the upper band \(W=+1\), whereas in the bottom band \(W=-1\), which means that there is amplification in one direction and attenuation in the other direction. To visualize the eigenmodes of the system, we introduce open boundaries in the \(y\)-direction and solve the eigenvalue problem. Fig. \ref{fig:fig4}(b) shows that in the presence of open boundaries, the wave is amplified in the bottom band (Fig. \ref{fig:fig4}(b)(i)) and attenuated in the top band (Fig. \ref{fig:fig4}(b)(ii)). Conversely, at the real-valued frequencies, we have bulk modes, as shown in Fig. \ref{fig:fig4}(b)(iii).

In conclusion, we have demonstrated that driven odd elasticity is possible in a mechanical structure using only passive elements, namely chiral gears and a square lattice metamaterial.  We achieve this leveraging periodic driving of chiral gears attached to the square metamaterial, where the chirality of the driven gears results in odd shear being developed when the metamaterial is subject to normal strains.  We demonstrate that the resulting elasticity tensor is odd and asymmetric, which enables non-conservative work to occur in a cycle of normal and shear strains, and that the odd metamaterial exhibits unique dynamical responses, including the well-known non-Hermitian skin effect.  We anticipate that the work presented here may motivate the development of other mechanisms in the future that can expand the scope and applicability of driven odd elasticity.

Both authors acknowledge the support of the AFOSR under award number FA9550-23-1-0299, while HSP also acknowledges NSF CMMI-2227474.

\bibliography{tis}% Produces the bibliography via BibTeX.

\end{document}